\documentclass[aps, prd, amsfonts, a4paper,floatfix,twocolumn]{revtex4}

\usepackage{graphicx}
\usepackage{amsmath,amsfonts}
\usepackage{epsfig,color}

\begin{document}
\newcommand{\be}{\begin{eqnarray}}
\newcommand{\ee}{\end{eqnarray}}
\newcommand\del{\partial}
\newcommand\nn{\nonumber}
\newcommand{\Tr}{{\rm Tr}}
\newcommand{\Str}{{\rm Trg}}
\newcommand{\mat}{\left ( \begin{array}{cc}}
\newcommand{\emat}{\end{array} \right )}
\newcommand{\vect}{\left ( \begin{array}{c}}
\newcommand{\evect}{\end{array} \right )}
\newcommand{\tr}{{\rm Tr}}
\newcommand{\hm}{\hat m}
\newcommand{\ha}{\hat a}
\newcommand{\hz}{\hat z}
\newcommand{\hx}{\hat x}
\newcommand{\hl}{\hat \lambda}
\newcommand{\tm}{\tilde{m}}
\newcommand{\ta}{\tilde{a}}
\newcommand{\tz}{\tilde{z}}
\newcommand{\tx}{\tilde{x}}
\definecolor{red}{rgb}{1.00, 0.00, 0.00}
\newcommand{\rd}{\color{red}}
\definecolor{blue}{rgb}{0.00, 0.00, 1.00}
\definecolor{green}{rgb}{0.10, 1.00, .10}
\newcommand{\blu}{\color{blue}}
\newcommand{\green}{\color{green}}



\title{New Ways to Determine Low-Energy Constants with Wilson Fermions}


\author{P.H. Damgaard}
\affiliation{Niels Bohr International Academy and Discovery Center, Niels Bohr Institute, University of Copenhagen,
  Blegdamsvej 17, DK-2100, Copenhagen {\O}, Denmark} 

\author{U.M. Heller}
\affiliation{American Physical Society, One Research Road, Ridge, NY 11961, USA}

\author{K. Splittorff}
\affiliation{Discovery Center, Niels Bohr Institute, University of Copenhagen, Blegdamsvej 17, DK-2100, Copenhagen
  {\O}, Denmark}

\date   {\today}
\begin  {abstract}
We show how the leading physical and Wilson low-energy constants 
associated with Wilson fermions in lattice gauge theory can
be determined individually by using spectral information of the Wilson
Dirac operator with fixed index at finite volume. The methods are 
demonstrated in simulations with leading-order improved Wilson fermions. 
In addition to the expected suppression of the leading term in Wilson 
chiral perturbation theory we observe a substantial reduction also of the 
higher-order Wilson low-energy constants.
\end{abstract}
\maketitle

\section{Introduction}
\label{sec:Intro}
One of the challenges facing lattice gauge theory simulations
near the chiral limit is how to disentangle ultraviolet lattice 
artifacts from chiral behavior. As simulations with Wilson
fermions approach the physical point of almost massless $u$ and $d$
quarks \cite{Luscher,Baron,Ishikawa:2007nn,Aoki:2009ix,Durr:2010aw}, 
this obstruction needs to be overcome. The traditional
method is to proceed through the Symanzik effective continuum
field theory which includes a series expansion in the 
lattice spacing $a$. Associated with each power of $a$ is
a set of higher-dimensional operators whose transformation
properties under chiral rotations determine the effective
low-energy field theory, Wilson chiral perturbation theory.
This program has been carried out to high order \cite{Sharpe}.
For a nice review, see, $e.g.$, Ref. \cite{Golterman}.

From the numerical point of view, the introduction of a whole
new set of Wilson low-energy constants in addition to the physical
low-energy parameters of QCD is, however, a challenge. 
One has to deal with several new fitting parameters,
and hence the need to fit more observables. Moreover, for typical
observables in the so-called $p$-regime of Wilson chiral
perturbation theory one finds that observables tend to depend
on particular combinations of the Wilson low-energy constants,
not directly on their individual values.

In this paper we wish to advocate a new strategy towards
determining the leading physical and Wilson low-energy constants which
is based on the new analytical results for spectral correlators 
of the Wilson Dirac operator \cite{DSV,ADSV,AN,KVZ,SV,KSV,Asger,Mario}. 
The idea is very simple, and relies on using the wealth of detailed
information contained in the spectral density and spectral
correlation functions of the Wilson Dirac operator $D_W$ and
its Hermitian counterpart $D_5$. Here, as in earlier work
on how to determine QCD low-energy constants based on the
Dirac operator spectrum \cite{Fpi}, the fact that one has a new finite-volume
scaling regime plays a crucial role in increasing the
accuracy of the method. In particular, we shall 
demonstrate how the chiral condensate, the quark mass and 
the Wilson low-energy constants $W_6$, $W_7$ and $W_8$ can be 
obtained from the Wilson Dirac spectrum. It is central for 
this method that one 
can divide lattice configurations into sectors of fixed index $\nu$, 
as determined by the spectral flow \cite{Heller0}. 
When combined with more traditional approaches from determining 
combinations of Wilson low-energy constants in the
$p$-regime by measuring, $e.g.$, small differences in
the charged and neutral pion masses \cite{Scorzato,ETMC} or from
other combinations of correlation functions \cite{Sharpe1},
one has a series of strong consistency checks on the
obtained values.

Perturbative results for the effect of $O(a^2)$ terms
on the spectrum have been compared to numerical simulations and 
used to determine Wilson low-energy constants in 
\cite{Necco:2011vx}. There have also been two initial studies of 
the detailed analytical predictions for the Wilson Dirac operator 
spectrum \cite{DHS,DWW}, results of which already looked promising. 
Here we explore the effect of introducing a simple ${\cal O}(a)$
clover improvement \cite{SW}, using the coefficient given by 
leading-order weak-coupling perturbation theory. As we shall
demonstrate below, one of the unexpected consequences of this
simple ${\cal O}(a)$ improvement is that also ${\cal O}(a^2)$
low-energy constants of Wilson chiral perturbation theory are 
substantially reduced. Moreover, certain 
asymmetries in the data of Refs. \cite{DHS,DWW} that clearly
could not be explained by Wilson chiral perturbation theory
up to, and including, ${\cal O}(a^2)$ effects are also 
substantially reduced by this clover improvement. This shows
that simple clover improvement even reduces the terms of order 
${\cal O}(am)$, where $m$ is the quark mass. 
This is good news for lattice simulations, and a
result that could not have been anticipated {\em a priori}.
All numerical data in this paper refer to the quenched approximation.
It would obviously be of interest to have a confirmation
of our results with the fermion determinant included.

Our paper is organized as follows. In the next section we briefly
review the theoretical setting and introduce our lattice set-up.
In section III we present our new numerical results, and explain
how different observables can be used to probe individual
Wilson low-energy coefficients. We end in section IV with 
conclusions and an outlook for future work.

\section{The Theoretical Framework}

The essential input that allows us to compute the effects of 
${\cal O}(a^2)$ on the Wilson Dirac operator spectrum is the
consistency, shown by Sharpe \cite{Sharpe2}, of the method 
up to contact terms of order $am$. 
As in in Refs. \cite{DSV,ADSV}, we use the $\epsilon$-regime counting
where $m \sim \epsilon^4$ and $a \sim \epsilon^2$, keeping terms up to
and including ${\cal O}(\epsilon^4)$ (the different possible counting
rules have been discussed in Ref. \cite{Shindler}). In the language of
finite-size scaling, it means that we are considering a regime in which
$$
m V, \quad z V, \quad a^2 V
$$
are kept fixed. Here $z$ is a source of the pseudoscalar
density $\bar{\psi}\gamma_5\psi$. 

Using the convention of Refs. 
\cite{DSV,ADSV}, we write the effective Lagrangian up to
that order as
\be
{\cal L}(U) & = &
\frac 12 (m+z) \Sigma { \rm Tr} U
+\frac 12 (m-z)\Sigma{ \rm Tr} U^\dagger \cr
&-& a^2W_6[{\rm Tr}\left(U+U^\dagger\right)]^2
     \!-\! a^2W_7[{\rm Tr}\left(U-U^\dagger\right)]^2 \nn \\
&-& a^2 W_8{\rm   Tr}(U^2+{U^\dagger}^2) , \label{L}
\ee
and we define the partition function in sectors of fixed
index $\nu$ by 
\be 
Z_{N_f}^\nu(m,z;a) & = & \int_{U(N_f)} d U \ {\det}^\nu U
 \ e^{-V {\cal L}(U)}.
\label{Zfull}
\ee
The quenched or partially quenched limits are defined in
the usual way. Operationally, the separation into sectors
of fixed index $\nu$ as given above corresponds
to a separation based on the spectral flow of the Wilson
Dirac operator \cite{ADSV}. This gives a double motivation
for following the spectral flow numerically, since it 
also provides us with the real eigenvalues of the Wilson
Dirac operator $D_W$ (see below).

We denote the usual Hermitian Wilson Dirac operator by
$D_5$:
\be
D_5 \equiv \gamma_5 (D_W +  m) ~.
\ee
The source $z$ introduced above couples to $\bar{\psi}\gamma^5\psi$. 
It is useful because the fermion determinant of the QCD path
integral is then, up to a sign, given by
\be
\det(\gamma_5(D_W + m)+z) = \det(D_5+z).
\ee
This prompts one to consider
the spectral resolvent of the Hermitian Wilson Dirac operator $D_5$,
\be\label{Gdef}
G(z) \equiv  \left \langle   \tr \left(\frac{1}{D_5+z - i\epsilon}\right)\right\rangle =
\left \langle \sum_k \frac 1{\lambda^5_k +z - i\epsilon}\right \rangle,
\ee
where $\lambda^5_k$ are the eigenvalues of $D_5$. From this, the density 
of eigenvalues of $D_5$ follows:
\be
\rho_5(\lambda^5) = \frac{1}{\pi}{\rm Im}\left.[G(-\lambda^5)]\right|_{\epsilon \to 0+}.
\label{rho5def}
\ee

Let us now briefly review some basic facts about the eigenvalues of
$D_W$ and $D_5$ \cite{ADSV}, many of which follow directly from the
early paper \cite{Itoh}. As is well known, eigenvalues of $D_W$ are 
either real or come in complex conjugate pairs. The real
eigenvalues of $D_W$ play particularly important roles, since
they provide a definition of gauge field topology as the lattice
spacing $a$ is taken to zero. Even at nonzero lattice spacing
these real modes are special. It follows from the definition of
$D_5$ that its eigenvalues $\lambda^5$ are functions of the mass $m$:
$\lambda^5 = \lambda^5(m)$. Tuning $m$ to a value $m_c$ at which a zero 
eigenvalue occurs, $\lambda^5(m_c) =0$, is particularly
interesting since
\be
D_5\phi ~=~ 0 ~~~ \Rightarrow~~~~ D_W\phi ~=~ -m_c\phi ~.
\ee 
This shows that finding a zero of $\lambda^5(m)$ corresponds to
identifying a real mode of $D_W$. It is straightforward to see that
the argument runs both ways, so that a real eigenvalue of $D_W$
also corresponds to a zero of $\lambda^5(m)$. Moreover, in a compact
notation, one can show that for given eigenstates $|j\rangle$ of
real modes of $D_W$ \cite{Itoh,ADSV},
\be
\left.\frac{d\lambda_j^5(m)}{dm}\right|_{m=m_c} ~=~ 
\langle j|\gamma_5|j \rangle
\ee
so that the slope at a crossing of the spectral flow is given
by the chirality of the state $|j\rangle$. As the continuum is
approached, this chirality goes to $\pm 1$. At any finite lattice
spacing the chirality vanishes identically for all nonreal modes 
\cite{Itoh}.

The chirality of the real modes is also directly related to the
index $\nu$ described above since around the
physical branch \cite{ADSV},
\be
\sum_{real}{\rm sign}\langle j|\gamma_5|j\rangle ~=~ \nu ~,
\ee
a relation which displays the topological nature of the index:
the number of zero-crossings, counted with signs. As the lattice
spacing $a$ is reduced the number of multiple crossings near
the physical branch of the spectrum goes
to zero. On typical configurations in the present study, the
definition of $\nu$ given by the spectral flow as shown above
agreed with the (improved) naive topological charge of the so-called
``Boulder method'' \cite{Boulder_Q}, which we applied after six HYP
smearings to the gauge fields. The few cases were the two determinations
differed were usually associated with multiple crossings, with the
last one at a fairly large $\lambda^W_{real}$.

We now remind the reader of a few basic facts about the spectrum
of the Hermitian Wilson Dirac operator \cite{Heller0}. In the continuum, 
the spectrum of the Dirac operator $D$ is {\em chiral}: for every nonzero 
eigenvalue $i\lambda$ there is a matching eigenvalue $-i\lambda$. This holds
configuration by configuration, and is a simple consequence
of the $\gamma$-matrix identity $\{\gamma^{\mu},\gamma^5\} = 0$.
The corresponding Hermitian Dirac operator $\gamma^5(D+m)$
is also chirally symmetric:
\be
\gamma^5(D+m)\phi_{\pm} ~=~ \pm\sqrt{\lambda^2 + m^2}\phi_{\pm}
\ee
and the pairing is preserved: for every eigenvalue pair $\pm i\lambda$
of $D$ there is a pair of eigenvalues $\pm\sqrt{\lambda^2 + m^2}$.
For the zero modes of $D$, where there is no chiral pairing,
there is also no pairing of eigenvalues of $\gamma^5(D+m)$:
these eigenvalues become either $+m$ or $-m$, depending on
their chiralities. The zero modes of $D$ are chiral eigenstates.
For the Wilson Dirac operator $D_W$ these properties are violated. 
In particular, the spectrum
of $D_W$ is in the complex plane and it is not chirally paired.
Likewise, the spectrum of $D_5 = \gamma^5(D_W+m)$ is not symmetric
configuration by configuration, although the density $\rho_5(\lambda^5)$
is symmetric in the $\nu=0$ sector.

Analytical predictions for the quenched spectrum of $D_5$ have been 
given in Refs.~\cite{DSV,ADSV}, although focus there was on the
simplifying case when only $W_8$ is considered nonvanishing.
General analytical expressions including $W_6, W_7$
and $W_8$ were also provided in those references. The constants 
$W_6, W_7$ and $W_8$ have fixed signs \cite{DSV,ADSV,Sharpe1,KSV}: Only for  
$W_6, W_7<0$ and $W_8>0$ is Wilson chiral perturbation theory the 
effective theory of lattice QCD with a $\gamma_5$-Hermitian Wilson Dirac 
operator.

For the clover-improved case studied here, we shall find that
$a\sqrt{|W_i| V} \ll 1$. In that case $\rho_5$ depends only on the
combination $|W_6|+|W_7|$ rather than on $W_6$ and $W_7$ separately
\cite{ADSV}. We shall therefore keep $W_7=0$ throughout the rest of
this paper, and show how we can disentangle the dependence on $W_6$
from that on $W_8$.

\section{Simulations and Numerical Results}

\begin{figure}[t!]
\includegraphics[width=8cm,angle=0]{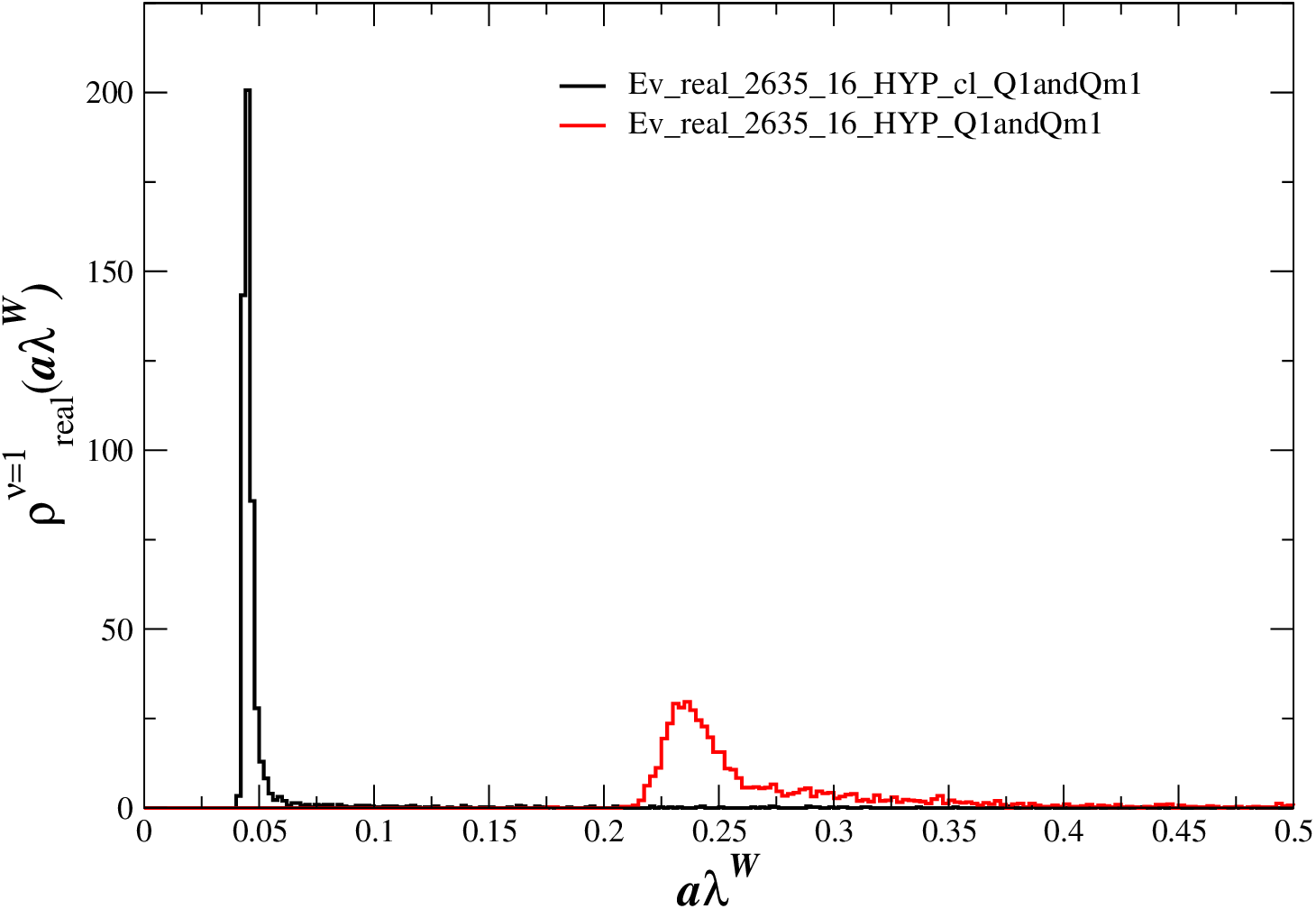}
\caption{\label{fig:rhoRlat} The effect of clover improvement on the density of the real eigenvalues of $D_W$. The black line is the clover improved data and the red line is the data before clover improvement. Not only does the peak shift towards the origin, the clover term also makes it much more narrow and symmetric.}
\end{figure}

We use two of the gauge field ensembles already considered in our previous
work \cite{DHS}. Both were generated with the Iwasaki gauge action
\cite{Iwasaki} and have a lattice size of $L=1.5$ fm. Some further properties
are given in Table~\ref{tab:configs}, including the lattice spacing
inferred from $r_0/a$ values from \cite{Iwasaki_r0.a,Iwasaki_r0.b}
using $r_0=0.5$ fm to set the physical scale, and the number of
configurations analyzed with $|\nu|=0$, 1, and 2. We have applied one HYP
smearing \cite{HYP} to the gauge fields before constructing the
clover Dirac operator with tree-level clover coefficient $c_{sw}=1$.
It is known \cite{HYP_clov} that the tree-level clover coefficient is
fairly close to a nonperturbatively improved value when using HYP
smeared gauge fields.

\begin{table}
\begin{tabular}{|l|c|c|l|r|r|r|}
\hline
$\beta_{Iw}$ & size & $a$ [fm] & $am_0$ & $\nu=0$ & $|\nu|=1$ & $|\nu|=2$ \\
\hline
2.635 & $16^4$ & 0.093 & -0.03  & 1276 & 2257 & 1518 \\
2.79  & $20^4$ & 0.075 & -0.027 & 1202 & 2128 & 1408 \\
\hline
\end{tabular}
\caption{Ensemble of pure gauge configurations considered, generated with
the Iwasaki gauge action. Listed are $\beta_{Iw}$, the size in lattice
units, the lattice spacing, the bare mass in the clover Dirac operator
and the number of configurations in the sectors with $|\nu|=0$, 1, and 2.}
\label{tab:configs}
\end{table}

We now turn to an analysis of our lattice data. We start by 
numerical results obtained on the $16^4$ lattice. As a first
test of the impact of clover improvement, we show in Fig.~\ref{fig:rhoRlat}
the density of real modes of $D_W$ in the $\nu=1$ sector. 
On the same plot is shown earlier 
data from Ref.~\cite{DHS} {\em without} clover improvement.
It is evident that clover improvement as expected shifts the 
position of the peak 
towards the origin. Moreover, it is also clear that the clover 
improvement has reduced the width of the peak substantially. In the 
analytic computation within Wilson chiral perturbation theory 
\cite{DSV,ADSV} this width is determined by the terms of order $a^2$. 
Hence we conclude that the clover term also has a positive effect on the 
order $a^2$ terms. Finally, the fact that the peak has become more 
symmetric must necessarily be due to an improvement of the terms 
of order ${\cal O}(am)\sim\epsilon^6$ and higher. As our analytical results 
extend only to order $\epsilon^4$  we unfortunately cannot quantify 
this suppression of the asymmetry further. Overall this seems to 
show that clover improvement works better than we had reasons to expect, 
see also \cite{Anna}. 
We now turn to the effect of clover improvement on the spectrum
of $D_5$. Since in the small-$a$ limit the real eigenvalues
of $D_W$ map directly to the peak at $\lambda^5 = m$, we
certainly expect improvement here as well. 

\begin{figure}[t!]
\includegraphics[width=8cm,angle=0]{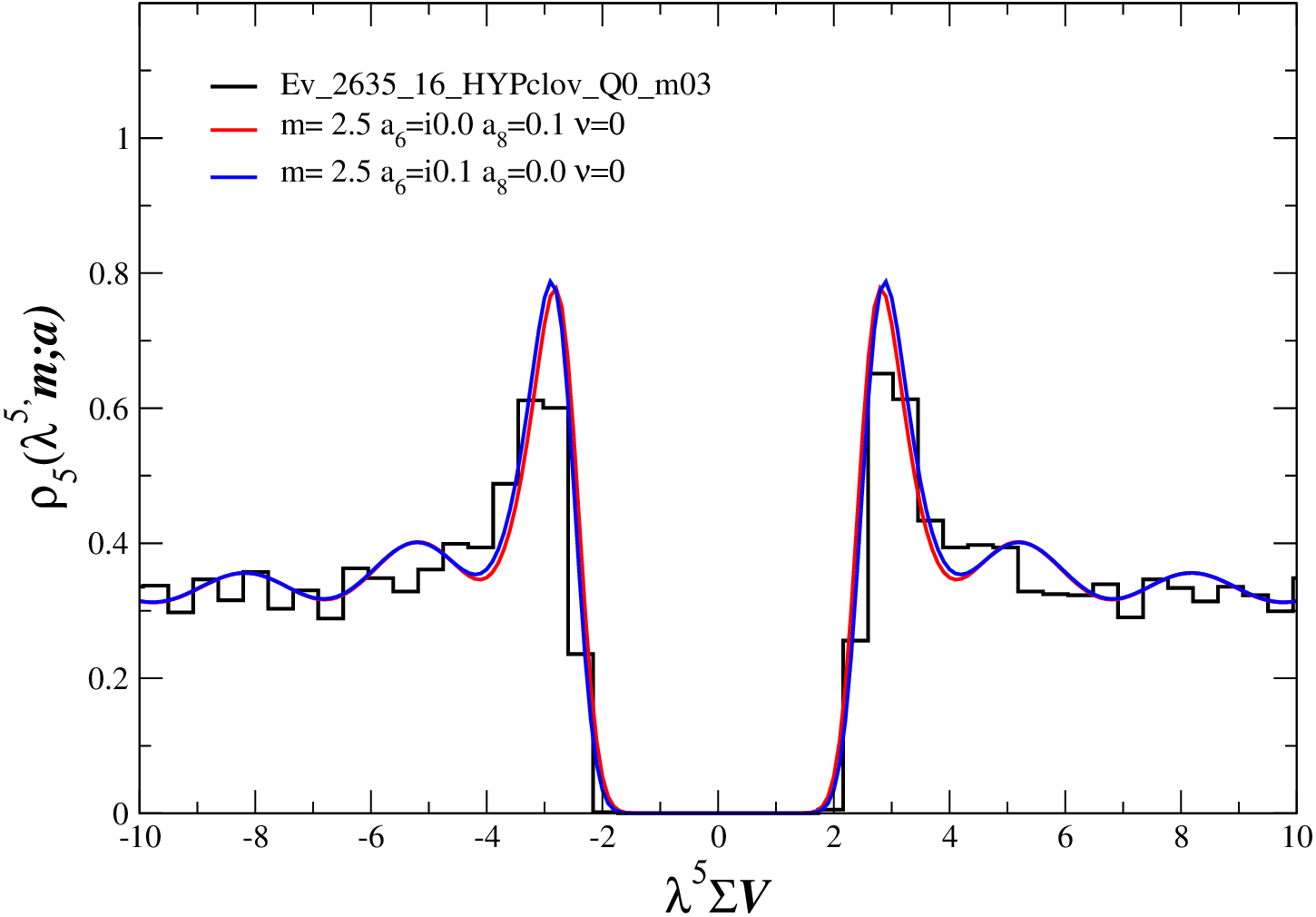}
\vspace{10mm}
\vfill
\includegraphics[width=8cm,angle=0]{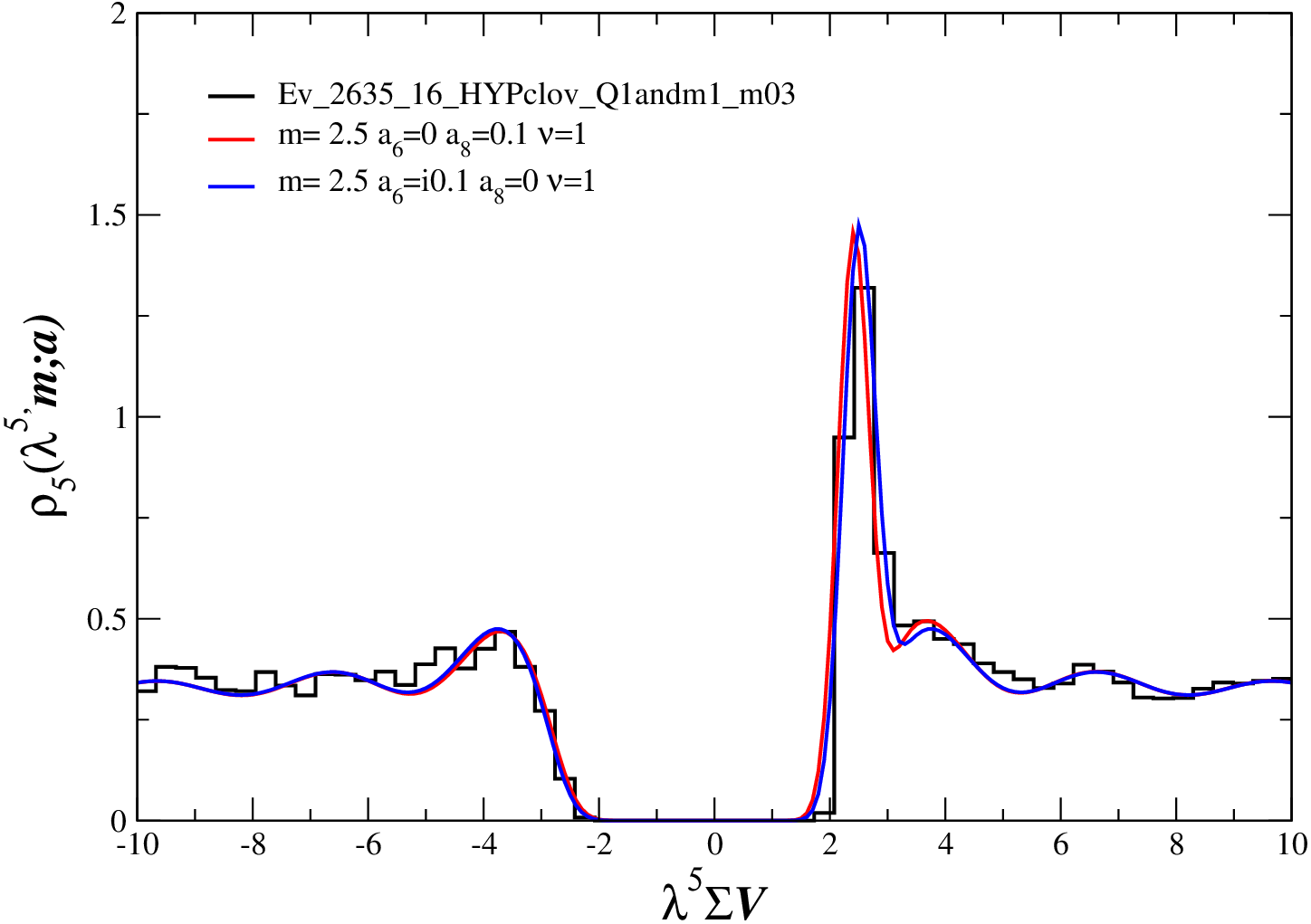}
\vspace{10mm}
\vfill
\includegraphics[width=8cm,angle=0]{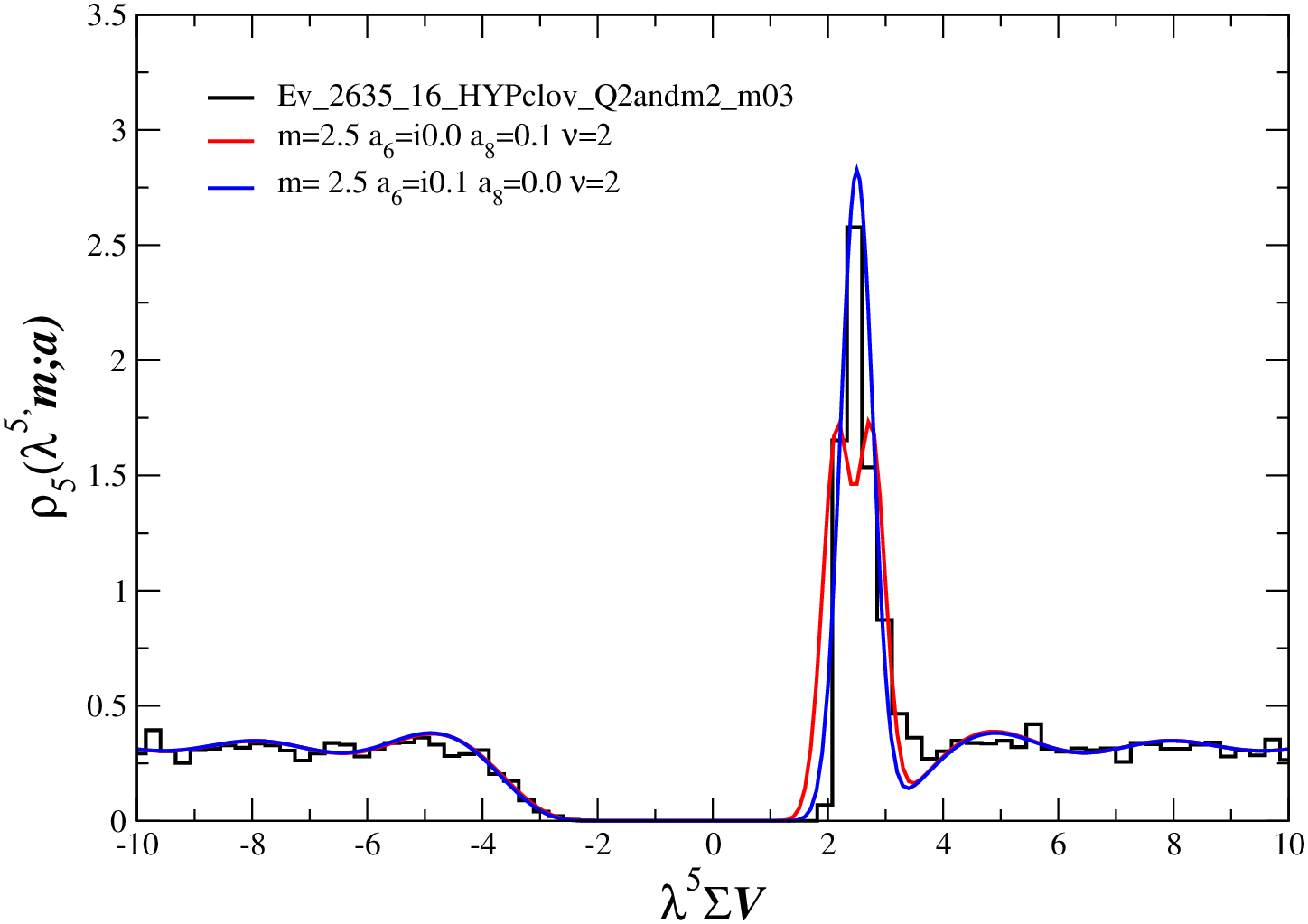}
\caption{\label{fig:rho5_16i4} The spectral density of the Hermitian 
Wilson Dirac operator on the $16^4$ lattices.
The numerically determined eigenvalues have been rescaled with $\Sigma V/a=173$
for the comparison to the analytic results.
The $|\nu|=2$ data in 
the lower panel allow us to separate the effect of $W_6$
from that of $W_8$.}
\end{figure}

The histograms in Fig.~\ref{fig:rho5_16i4} give the eigenvalue 
density of the Hermitian Wilson Dirac operator in the $\nu=0,1$ 
and $2$ sectors as obtained from the $16^4$ lattices. Displayed 
also are two fits of the analytic curves obtained from Wilson 
chiral perturbation theory. 
With the additional Wilson low-energy constants there are in total 5 
parameters to fit ($\Sigma V/a$, $m\Sigma V$, $W_6a^2V$, $W_7a^2V$ and 
$W_8a^2V$). Fortunately the analytic results of \cite{DHS,ADSV}
give us a series of insights that allow us to address these constants 
in turn. 
First of all, as mentioned above, the Wilson low-energy constants 
$W_i$ have fixed signs $W_6<0$, $W_7<0$ and $W_8>0$ 
\cite{DSV,ADSV,Sharpe1,KSV}.  
Moreover, \cite{ADSV,AN} in the limit where $a\sqrt{|W_i| V}\ll1$ 
there is a factorization of the eigenmodes of $D_5$, while those in 
the index peak have a specific dependence on the $W_i$'s the 
rest of the eigenvalue density is not affected to leading order in 
$a\sqrt{|W_i| V}$, see Fig.~\ref{fig:a0VSfit} below for an illustration. 
This has the following most useful consequences: 
\vspace{3mm}

{\sl 1)} The eigenvalue density on the opposite side of the index peak is 
almost continuum-like. 

{\sl 2)} The smallest eigenvalues on this side of 
the origin are located very close to $-m$ and thus offer a clean way 
to extract the quark mass parameter.   
 
{\sl 3)} The value of $\Sigma$ can be estimated by scaling the 
$\lambda^5$ axis and $m$ until the data on this side of the gap match 
the continuum predictions.   

{\sl 4)} The values of the $W_i$'s are finally obtained from their 
effect on the index peak (see below). 
\vspace{3mm}

The fits were made, using the above insights, to the $|\nu|=1$ data and have 
$a\sqrt{-W_6 V}=0.1$ and $W_8=0$ (blue line)  
respectively $W_6=0$ and $a\sqrt{W_8 V}=0.1$ (red line). 
In both cases $\Sigma V/a=173$ and $m\Sigma V=2.5$.

As can be seen from the middle panel of Fig.~\ref{fig:rho5_16i4} 
the $|\nu|=1$ data work well both with $W_6$ alone (blue curve) 
and with $W_8$ alone (red curve).

Having fixed the values of the low-energy constants by the 
$|\nu|=1$ data we can now check the two corresponding predictions 
from Wilson chiral perturbation theory against the $\nu=0$ and 
$|\nu|=2$ data. In the top panel for $\nu=0$ both predictions do well, 
however, for the $|\nu|=2$ data only the prediction from the fit 
with $W_6$ alone reproduces the peak structure of the would-be 
topological modes.  

\begin{figure}[t!]
\includegraphics[width=8cm,angle=0]{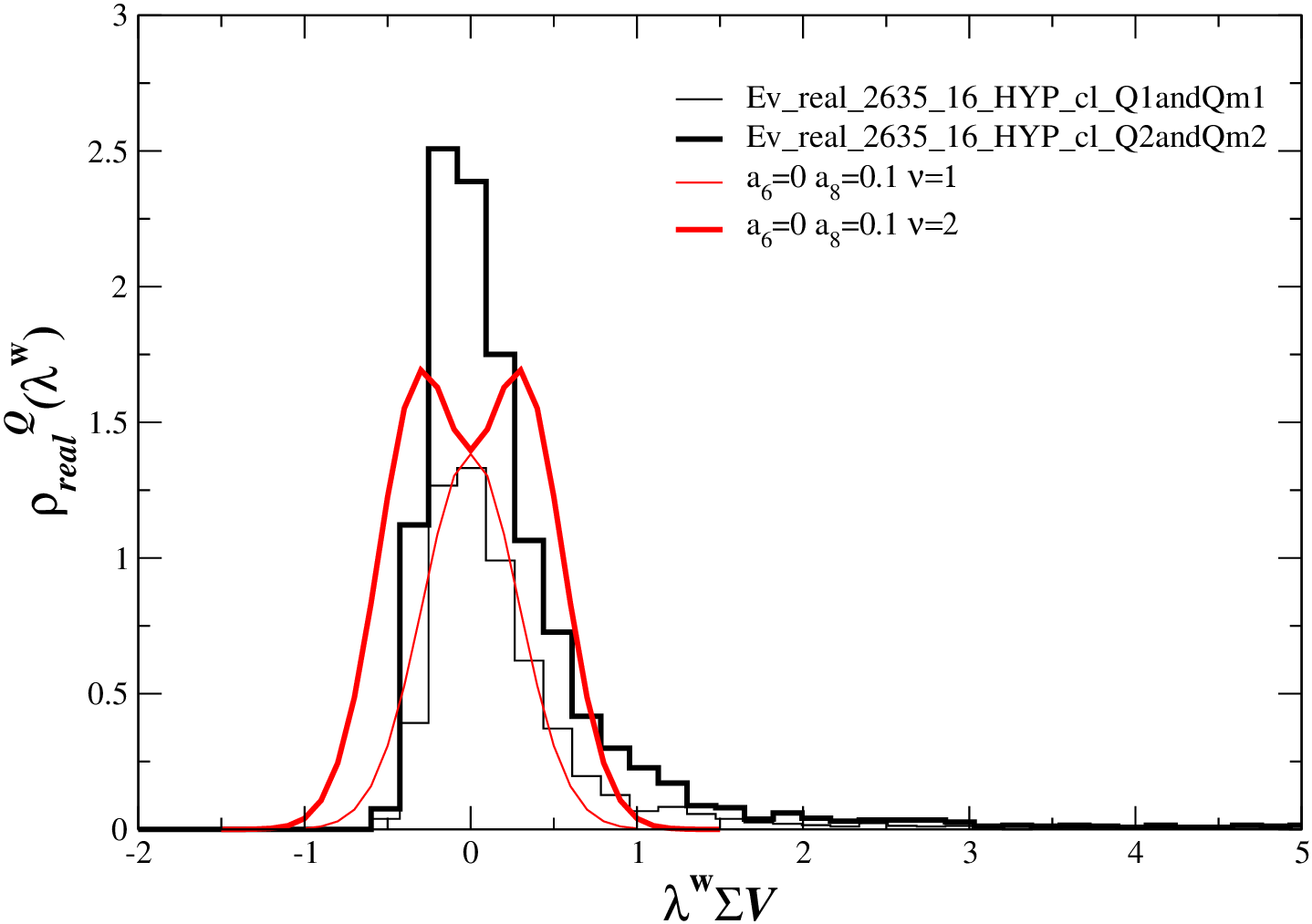}
\vspace{10mm}{}
\vfill
\includegraphics[width=8cm,angle=0]{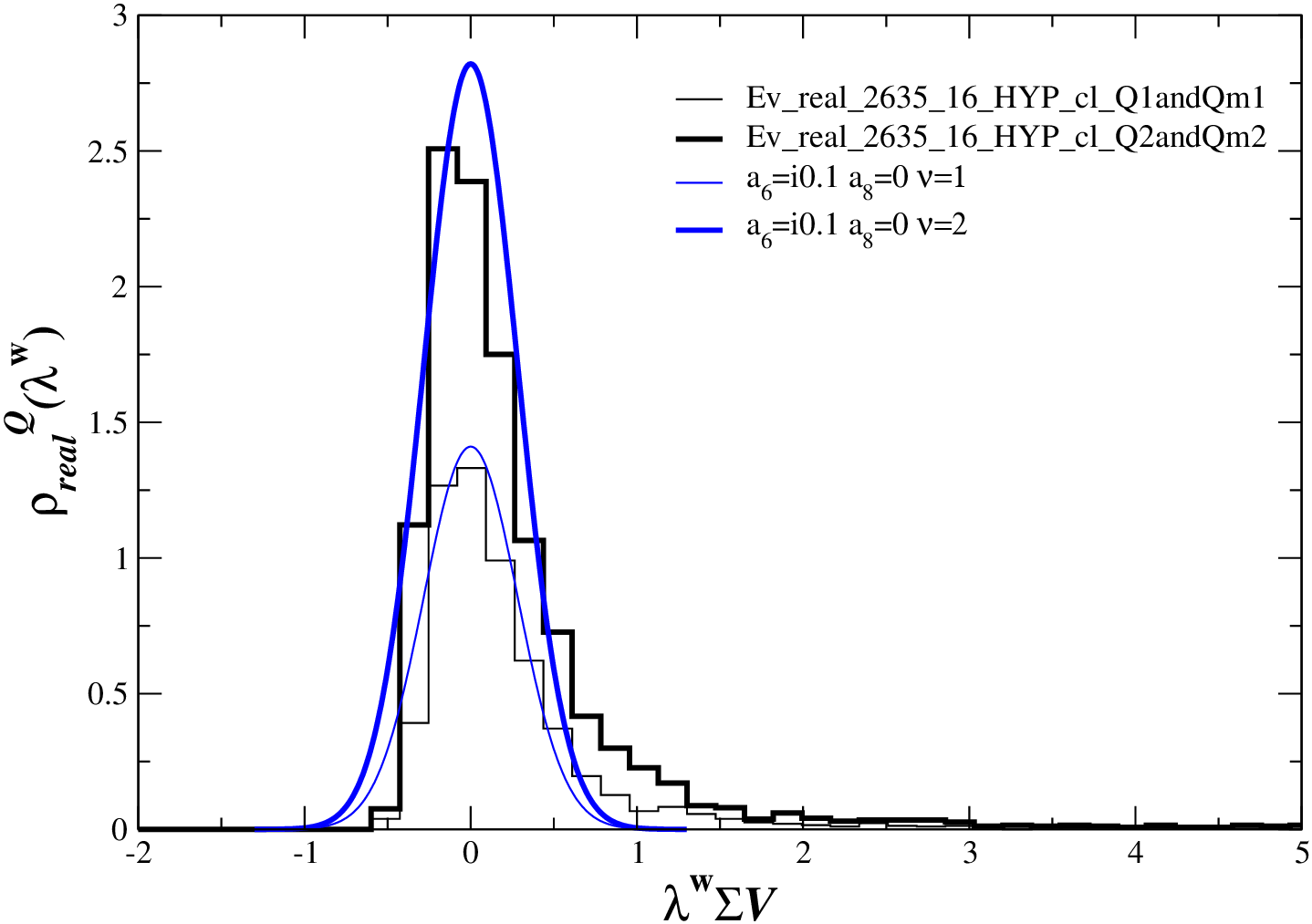}
\caption{\label{fig:rhoR16i4} The density of the real eigenvalues of $D_W$,
rescaled with $\Sigma V/a=173$. 
Clearly the prediction with $W_8=0$ (blue lines lower panel) gives the 
better description of the $|\nu|=2$ data.}
\end{figure}

\begin{figure}[t!]
\includegraphics[width=8cm,angle=0]{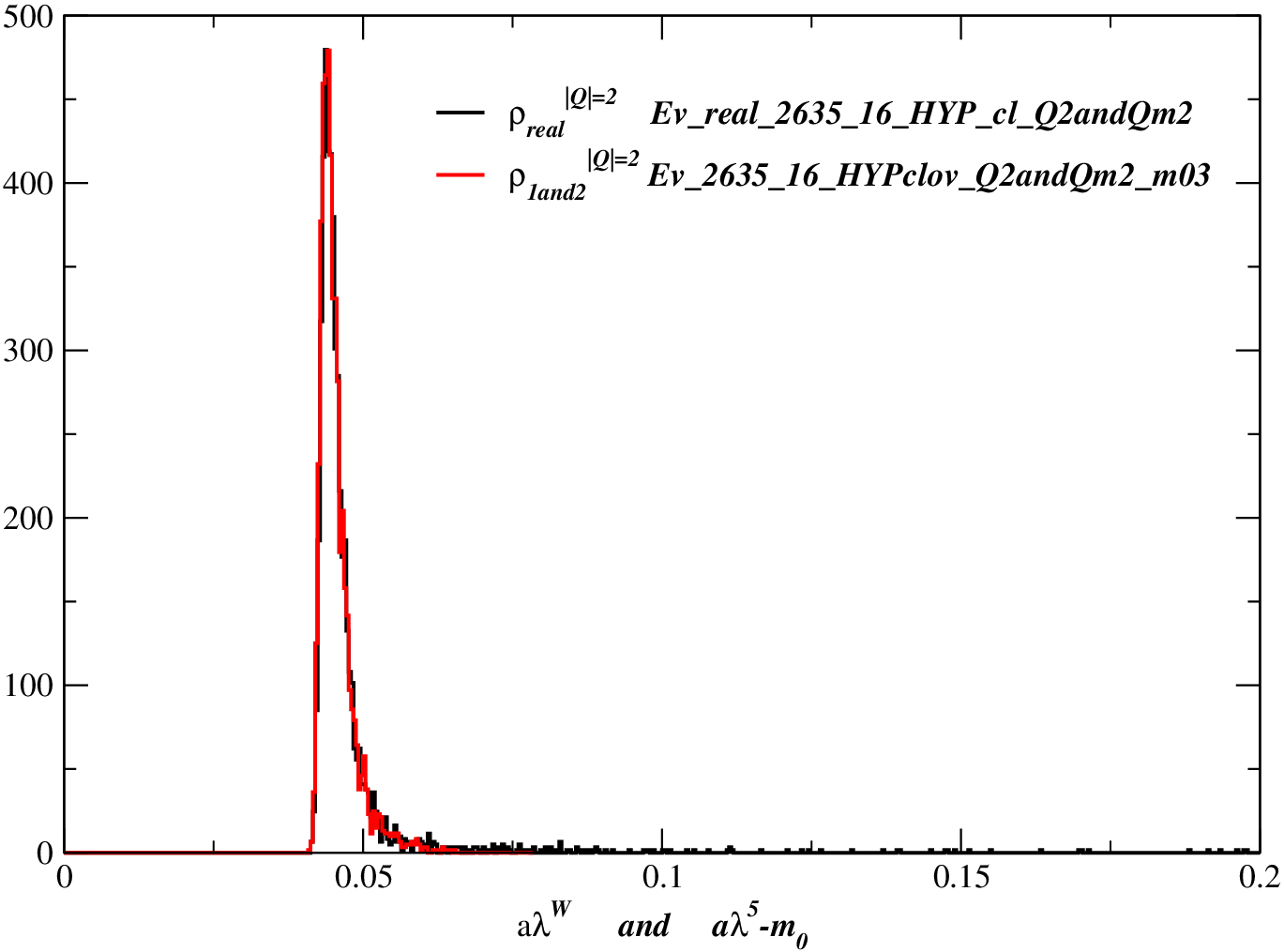}
\caption{\label{fig:rhoRVS12} The density of the real eigenvalues of
  $D_W$ for $|\nu|=2$ on the $16^4$ lattice plotted together with the
  accumulated density of the first two positive eigenvalues of $D_5$ 
on the same configurations shifted by the bare mass. The excellent match
  between the two demonstrates that the corresponding eigenvectors are
  almost chiral.}
\end{figure}

\begin{figure}[t!]
\includegraphics[width=8cm,angle=0]{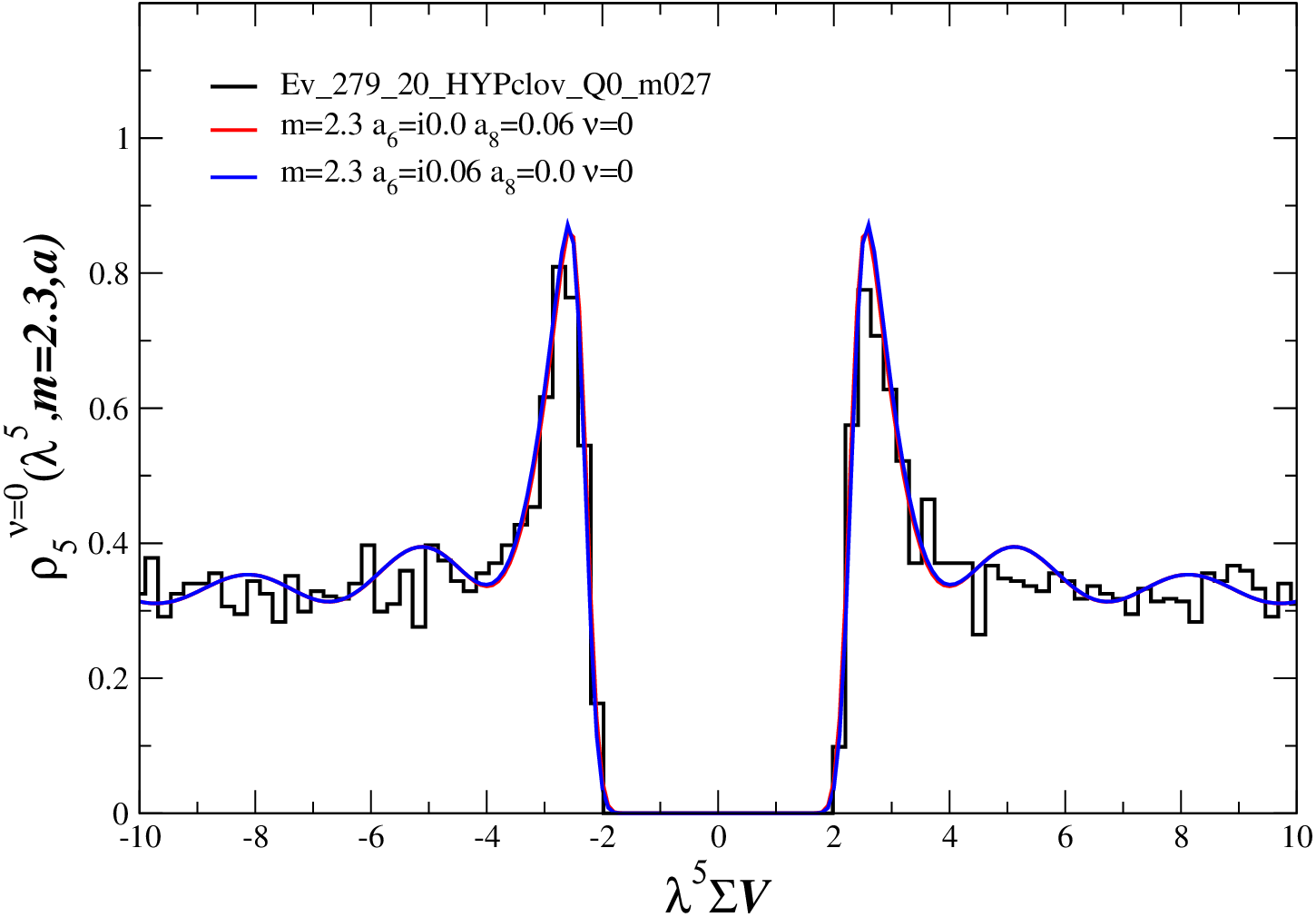}
\vspace{10mm}
\vfill
\includegraphics[width=8cm,angle=0]{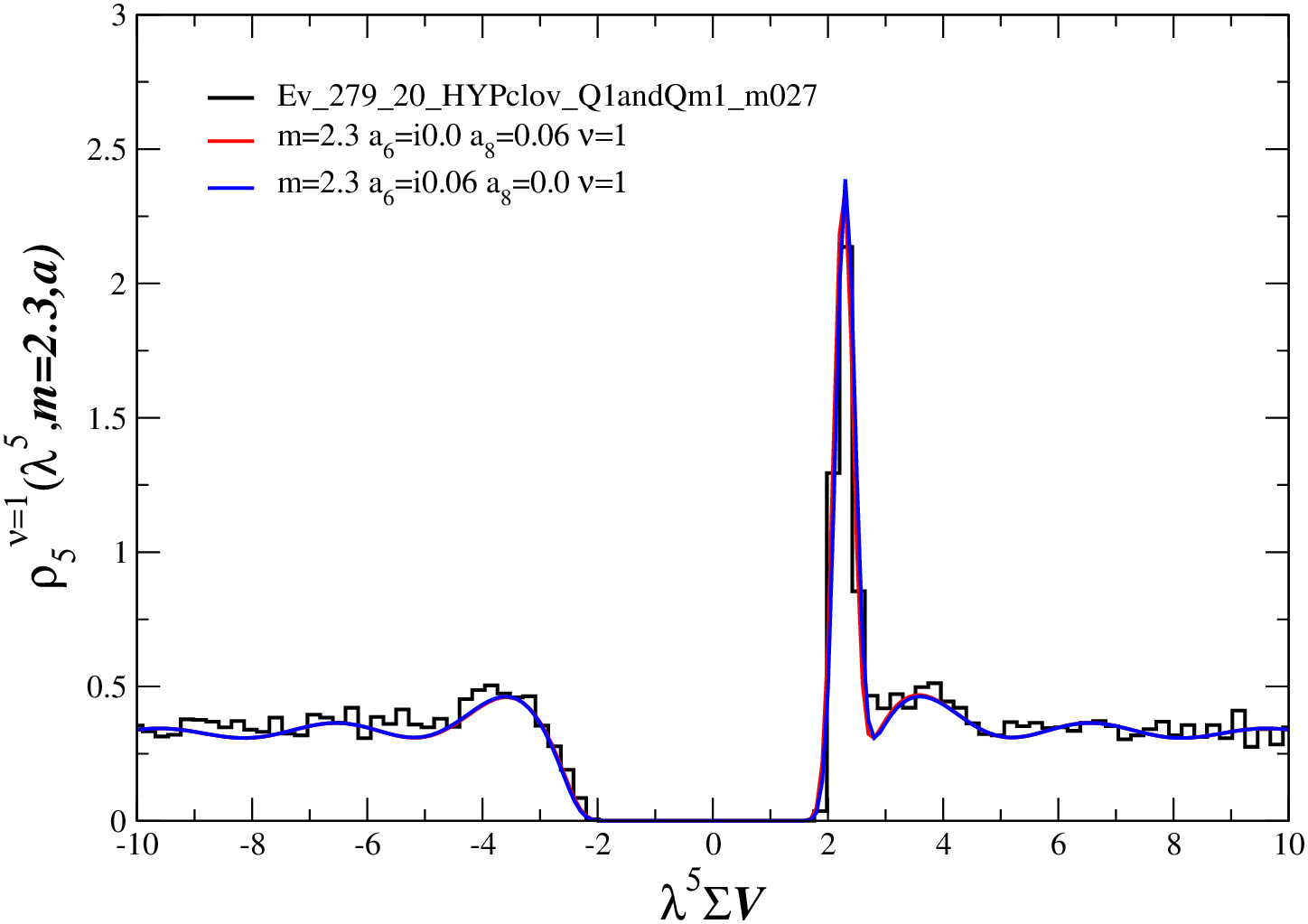}
\vspace{10mm}
\vfill
\includegraphics[width=8cm,angle=0]{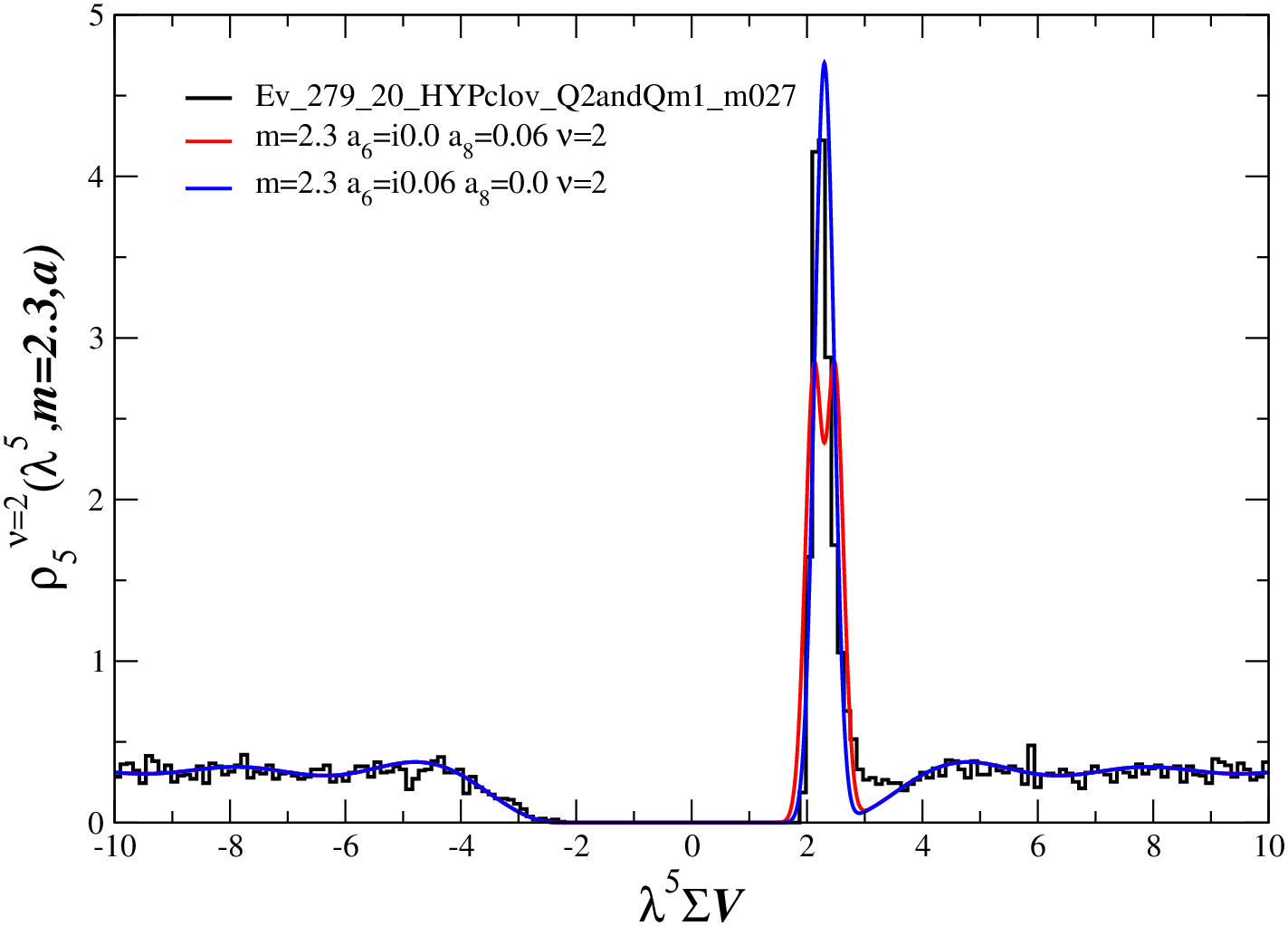}
\caption{\label{fig:rho5_20i4} The spectral density of the Hermitian Wilson Dirac operator on the $20^4$ lattices. The eigenvalues have been rescaled
with $\Sigma V/a=220$. In blue and red are two fits (see the main text) 
to the $|\nu|=1$ data in the middle panel. In the top and lower panel the 
predictions obtained from these fits are plotted against the data.}
\end{figure}

In Fig.~\ref{fig:rhoR16i4} we plot the analytic predictions (the values 
of the low-energy constants still fixed as above) for 
the real eigenvalues of the Wilson Dirac operator against the real 
eigenvalues of the $|\nu|=1$ and $|\nu|=2$ sectors. 
In the top panel the red lines display the predictions 
corresponding to the fit with $W_8$ alone. The predictions 
from the fit with $W_6$ alone are given in the lower panel 
(blue lines). Again the $|\nu|=2$ data for the real 
eigenvalues clearly favors the prediction with $W_6$ alone.

The reason why $W_6$ and $W_8$ have such a different effect on the 
analytic prediction with $|\nu|=2$ is the following: The $W_6$-term 
in the $\epsilon$-regime of Wilson chiral perturbation theory 
corresponds to a Gaussian fluctuating mass \cite{ADSV}. 
The $\delta$-peak from the continuum at $m$ is therefore smeared 
into a Gaussian with an amplitude that simply scales with $\nu$. 
In particular, $W_6$ does not induce eigenvalue repulsion between 
the real modes. On the contrary 
it was shown in \cite{ADSV,AN} that for small but nonzero 
$a^2W_8V$ the density 
of the real eigenvalues of $D_W$ takes the form of the $\nu\times\nu$ 
Gaussian unitary ensemble scaled by $4a^2W_8V$. The effect of $W_8$ is 
therefore to induce the eigenvalue repulsion familiar from random 
matrix theory between the real modes. In this way we get a clear 
distinction between the effect of $W_6$ and $W_8$. 

The fact that the real modes of $D_W$ and the nearly topological peak 
in the spectrum of $D_5$ both lead to the same conclusion is not 
accidental. In Fig.~\ref{fig:rhoRVS12} we plot the real modes of $D_W$ 
for $|\nu|=2$ together with the distribution of the first two positive 
eigenvalues of $D_5$ appropriately shifted by the bare mass. The 
essentially perfect match between the two distributions demonstrates 
that the corresponding eigenvectors are almost chiral as is expected 
for $|a^2W_iV|\ll1$, see figure 4 of \cite{ADSV} and the associated 
discussion.

Let us now consider the $20^4$ data set. 
In Fig.~\ref{fig:rho5_20i4} the eigenvalue density of the Hermitian 
Wilson Dirac operator is displayed for $\nu=0$ (top panel), $|\nu|=1$ 
(middle panel) and $|\nu|=2$ (lower panel). Again we have made two fits 
to the $|\nu|=1$ data one with $W_8=0$ (blue line) and one with 
$W_6=0$ (red line). The values in this case are 
$a\sqrt{-W_6 V}=0.06$ and $W_8=0$ (blue line)  
respectively $W_6=0$ and $a\sqrt{W_8 V}=0.06$ (red line) both with 
$\Sigma V/a=220$ and $m\Sigma V=2.3$. 

These values were determined according to the strategy outlined above. 
To illustrate the factorization of the behavior used we show in 
Fig.~\ref{fig:a0VSfit} the fit for $\nu=1$ and $W_8=0$ together 
with the continuum, $a=0$, curve at the same quark mass. As is clear 
the leading-order effect for $a\sqrt{|W_i|V}\ll1$ is on the index peak
only.   

The two fits both describe the 
$|\nu|=1$ data well and the prediction for $|\nu|=0$ also works 
nicely for both fits.
For $|\nu|=2$ the analytic predictions again differ in the region of the 
topological modes. Also here it is the prediction with $W_6$ alone 
that gives 
the only acceptable fit to the $|\nu|=2$ data. This again allows us to 
conclude that $W_6$ gives 
the dominant contribution to the discrepancy from the continuum.

\begin{figure}[t!]
\includegraphics[width=8cm,angle=0]{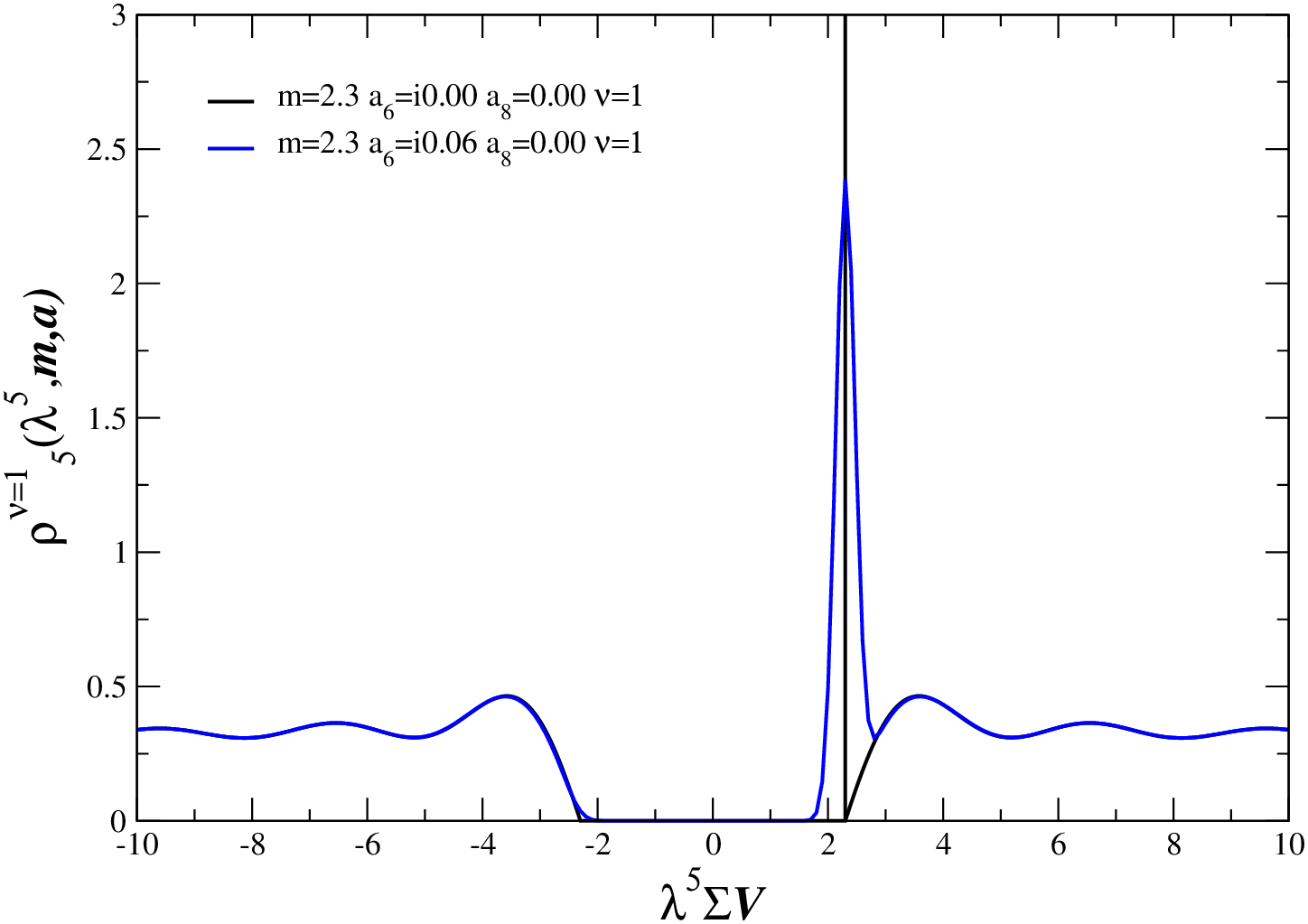}
\caption{\label{fig:a0VSfit} Comparison of the microscopic eigenvalue 
density of the Hermitian Wilson Dirac operator in the continuum 
$a=0$ (black curve) and with $a\sqrt{W_6V}=i0.06$, $W_8=0$ (blue curve).
The vertical black line indicates the position of the topological 
$\delta$-peak.  To leading order in $a\sqrt{|W_i|V}$ only this peak 
is affected by the discretization.}
\end{figure}

Finally, to summarize the effect of clover improvement: Before clover 
improvement \cite{DHS} we found that the values of $a\sqrt{|W_8|V}$ 
where roughly $0.35$ for both the $16^4$ and $20^4$ lattices. Here, 
after clover improvement, we find that $a\sqrt{|W_8|V}$ is consistent with zero and that $a\sqrt{|W_6|V}$ takes the values $0.10$ and $0.06$, respectively. 
Moreover, the asymmetry of the density of the real modes has been 
substantially reduced by the addition of the clover term. This 
indicates an improvement also at order ${\cal O}(am)$. While this 
unexpected order ${\cal O}(a^2)$ and ${\cal O}(am)$ improvement 
is an analytic challenge to understand, it is clearly 
good news for lattice QCD.

\section{Conclusions and Outlook}

We have shown how the microscopic eigenvalue density of the Wilson Dirac
operator can be used to determine both physical and lattice-artifact
low-energy constants from lattice QCD simulations. We have demonstrated
this with quenched clover-improved simulations. The method relies crucially
on our ability to divide the lattice configurations into sectors with
fixed index.

While initial numerical tests \cite{DHS,DWW} had already demonstrated
the feasibility of computing the microscopic spectrum of the Wilson Dirac
operator and comparing it to analytical predictions, there were aspects 
of the measured spectra that displayed
a systematic disagreement with the analytical predictions. In particular, 
the observed asymmetry in the spectrum of the real modes of $D_W$. This
asymmetry has been known to be present in data from very early on,
and yet it is a firm prediction of Wilson chiral perturbation theory
up to and including ${\cal O}(a^2)$ that the spectrum must be symmetric.
A quick analysis reveals that terms involving odd powers of $am$ in
the effective theory are needed to explain such an asymmetry. This
is beyond present analytical predictions. Here we have improved at 
order $a$ through use of the conventional 
clover term. Since the effective theory contains an arbitrary linear
shift in the mass term anyway, such an improvement is, in this context, 
not interesting in itself. But clearly clover improvement has an impact
on higher-order coefficients as well. We have found that clover
improvement in fact substantially reduces the higher-order Wilson
constants {\em and} makes the spectrum of real modes much more
symmetric, and hence in good agreement with analytical predictions.
Because of the close connection between real modes of
$D_W$ and the ``topological'' (threshold) eigenvalues of
$D_5$, this improves the eigenvalue spectrum of $D_5$ as well.  

Apart from demonstrating substantially better agreement between theory
and numerical data, we have taken the opportunity to explain new
ways to measure Wilson low-energy constants based on spectral
data of the Wilson Dirac operator. In particular, we have shown
how detailed understanding of how the different ${\cal O}(a^2)$ 
operators in the Wilson chiral Lagrangian influence the
spectrum can be used to isolate dependencies on the individual
low-energy constants. This should be very helpful in determining
the numerical values of the constants in dynamical simulations.
Crucial in this context is the separation of configurations
into sectors of fixed index $\nu$, the Wilson analogue of
a topological charge. In each sector there are definite predictions
with which to compare the data.

The present study should be extended to full-scale simulations
with dynamical fermions. Analytical predictions are already available 
for two light flavors \cite{SV}. This gives entirely new ways to 
measure the low-energy constants of Wilson chiral perturbation theory, 
that can be compared to and combined with alternative approaches 
using more conventional space-time dependent observables in the 
$p$-regime. Although computationally more complicated,
it would also be most interesting to carry out analogous studies of the
full complex spectrum of $D_W$. The analytical predictions are 
available in \cite{KVZ,KSV}. In particular, the realization of either 
the Aoki phase \cite{Aoki} or the Sharpe-Singleton scenario \cite{Sharpe}  
is closely linked to the complex spectrum of $D_W$ \cite{KSV}.

Finally, due to the unexpected higher order improvement caused by 
the clover term it would be 
interesting to test the effect of clover improvement also in twisted mass 
lattice QCD. As we have demonstrated here, the spectral density of the 
lattice Dirac operator is an efficient tool for such an analysis. The
necessary analytical predictions from Wilson chiral perturbation with 
a maximally twisted mass have recently been derived \cite{SVtwist}.

\vspace{0.3cm}
\noindent
{\bf Acknowledgments:}
We thank Jac Verbaarschot, Urs Wenger and Silvia Necco for discussions. 
PHD and KS would like to thank the CERN theory group for discussions and 
hospitality while this work was completed. 
The work of K.S. was supported by the {\sl Sapere Aude} program of
The Danish Council for Independent Research.


\end{document}